# A new high pressure form of Ba$_3$NiSb$_2$O$_9$

Céline Darie[a,*], Christophe Lepoittevin [a], Holger Klein [a], Stéphanie Kodjikian [a], Pierre Bordet [a], Claire V. Colin [a], Oleg I. Lebedev [c], Catherine Deudon [b], Christophe Payen [b]

[a] Institut Néel, Université Grenoble-Alpes, F-38042 Grenoble, France and Institut Néel, CNRS, F-38042 Grenoble, France

[b] Institut des Matériaux Jean Rouxel (IMN), UMR 6502, Université de Nantes, CNRS, F-44322 Nantes Cedex 3, France

[c] Laboratoire CRISMAT, ENSICAEN UMR6508, 6 Bd Maréchal Juin, Cedex4, Caen F-14050, France

## 1. Introduction

The Quantum Spin Liquid state was first proposed by Anderson in 1973 as the ground state for a system of spins on a triangular lattice that interact with their nearest neighbors via an antiferromagnetic interaction [1]. Quantum Spin Liquid materials have been the subject of many studies over the last decades [2]. Recently, candidates from the family of 6H-perovskites Ba$_3$MSb$_2$O$_9$ (M= divalent transition metal) have retained the attention and led to a few experimental studies [3, 4, 5, 6]. The case of Ba$_3$NiSb$_2$O$_9$ seems particularly interesting since it could represent an experimental proof of a Quantum Spin Liquid with spin S=1. This compound was first mentioned by Blasse et al. [7]. A first realization with structure solution was reported by Köhl *et al*. [8]. Since this work, three different structure types were observed and reported. All of them are built up by the stacking along the **c** axis of layers made of either NiO$_6$ and SbO$_6$ octahedra and differ by the way layers are stacked via either corner or face sharing of the octahedra. The first structure type was called either I [9] or 6H-A [10]. It consists of layers of isolated Sb$_2$O$_9$ dimers made of face-sharing SbO$_6$ octahedra linked by corner-sharing to layers of NiO$_6$ octahedra (fig. 1a). The space group is *P6$_3$/mmc* with parameters *a* = 5.8357(2)Å and *c* = 14.3956(4)Å [11]. The Ni$^{2+}$ cations occupy the 2a Wyckoff position and this site forms a planar triangular lattice perpendicular to the **c**-axis (fig. 1a and c). The second structure type called either II [9] or 6H-B [10] was reported in the case of application of a High Pressure High Temperature (HP-HT) treatment in the range 3 to 6 GPa at 600°C. With increasing pressure up to 9 GPa Cheng et al. [10] stabilized a new phase called 3C with a cubic structure (space group: *Fm3m*, *a* = 8.1552(2)Å for M=Ni). The structural arrangement of 6H-B was described in the *P6₃mc* space group and presents slightly smaller cell parameters: *a* = 5.7923(2)Å and *c* = 14.2922Å for M=Ni [10]. In that case the dimers made of face-sharing octahedra are occupied by both $M^{2+}$ and $Sb^{5+}$ cations and they are linked by corner-sharing to $SbO_6$ octahedron layers along the **c** axis (fig. 1b). For both structure types, the $Ba^{2+}$ cations are located in the voids generated by the stacking of the layers of octahedra and adopt a 12-fold coordination. As seen in fig. 1, there is for each structure type a definite stacking order between the $M^{2+}$ and $Sb^{5+}$ cations layers along the **c**-axis. If A(C) denotes the octahedron layers occupied by cations C, the stacking sequence of the 6H-B structure type would be -A($Sb^{5+}$)-A($Sb^{5+}$)=A($M^{2+}$)-A($Sb^{5+}$)-A($Sb^{5+}$)=A($M^{2+}$)-…, while that of the 6H-A would be -A($M^{2+}$)-A($Sb^{5+}$)=A($Sb^{5+}$)-A($M^{2+}$)-A($Sb^{5+}$)=A($Sb^{5+}$)-…, where the "–" sign represents corner sharing between layers, and the "=" sign represents face sharing. A striking difference between both stacking sequences is that for 6H-A, the face sharing octahedra dimers are occupied by the same $Sb^{5+}$ cations and the $M^{2+}$ cations occupy the single octahedron layers, while for phase 6H-B both types of cations occupy the dimers and the single octahedron layers are occupied by $Sb^{5+}$ cations. From the symmetry point of view, the mirror perpendicular to the **c**-axis and containing the common face between the octahedra forming the dimers is present in the 6H-A phase, but absent in the 6H-B phase, hence the symmetry lowering from *P6₃/mmc* to *P6₃mc*. It is interesting to note that both stacking sequences are compatible with the $6_3$ screw axis and **c** glide plane. However, a complete disorder between the cations forming the dimers would reestablish the higher *P6₃/mmc* symmetry.

We focus here on the form II or 6H-B of the $Ba_3NiSb_2O_9$ compound. In a previous study Cheng *et al*. [10] reported that this compound presents no magnetic order down to 0.35 K and suggested that it could be an experimental realization of a gapless quantum spin-liquid. In order to investigate this behavior by both NMR (nuclear magnetic resonance) and µSR [12] we synthesized this phase in HP-HT conditions. Preliminary characterizations of the resulting powder by X-ray diffraction seemed to indicate a possible disorder on the Ni/Sb site distribution. Indications for the existence of structural disorder in these phases have been already reported in the literature. Blasse *et al*. noticed weak intensities and broadening of specific reflections on X-Ray Powder Diffraction patterns of 6H-B $Ba_3NiSb_2O_9$, leading them to the conclusion that the long-range order was "difficult to attain" in the compound [7].

EXAFS experiments led Battle *et al*. to conclude to a "much less regular environment around the Ni$^{2+}$ cations" than mentioned in the previous works [9]. More recently, a Ni/Sb disordered distribution was reported on the cation sites in the related compound Sr$_3$NiSb$_2$O$_9$ [13]. For a correct interpretation of physical properties of our samples, it was mandatory to clarify the nature of the disorder and confirm (or not) the existence of Ni triangular planes in the structure. Therefore, we have undertaken a thorough investigation of the synthesis conditions and structural properties of 6H-B Ba$_3$NiSb$_2$O$_9$ This work using a combined approach coupling X-ray, neutrons and electron diffraction and microscopy, lead us to reconsider the crystal structure of 6H-B Ba$_3$NiSb$_2$O$_9$ powder samples obtained by HP-HT synthesis.

## 2. Experimental section

### 2.1. Synthesis

Polycrystalline samples of 6H-B Ba$_3$NiSb$_2$O$_9$ were synthesized by treatment under HP-HT of 6H-A Ba$_3$NiSb$_2$O$_9$. The ambient pressure 6H-A form was prepared by a conventional solid-state reaction. First NiSb$_2$O$_6$ was prepared by reaction of NiO (Strem, 99.99%) and Sb$_2$O$_5$ (Alfa, 99.998 %) in air at 1000 °C. Stoichiometric amounts of BaCO$_3$ (Aldrich, 99.98 %) and NiSb$_2$O$_6$ were then heated at 1000°C for 68 h and 1200 °C for 70h in air with an intermediate regrinding [14]. Samples of the 6H-A phase were characterized by powder X-ray diffraction (XRPD) on a Bruker D8 Advance instrument using monochromatic Cu-K α1 (λ= 1.540598 Å) radiation. Pure samples of 6H-A Ba$_3$NiSb$_2$O$_9$ were then packed into gold capsules and heated in either a belt or a Conac 28 anvil-type apparatus. The latter device allows elaborating a larger amount of powder (~ 350mg) in a single experiment. Several syntheses were performed with the belt anvil system in order to determine the best conditions for the syntheses. The explored condition range was between 2 and 4 GPa and 400°C to 680°C. The resulting phases were characterized by XRPD analysis.

We could observe several effects of either temperature and pressure: i) when the temperature is below 500°C no transformation occurred, the 6H-A phase was stable under high pressure, ii) the best results for synthesis of the 6H-B phase were obtained for a temperature of 620°C close to that reported by Chen (600°C) , iii)  for a pressure higher than

2.7 GPa we systematically obtained a mixture of the 6H-B and 3C forms. The amount of 3C form increased with pressure. It is interesting to note that the synthesis of pure 3C phase is reported with a much higher pressure of 9 GPa [10]. We noticed that decreasing the pressure allowed us to avoid the formation of the 3C form, undesirable in our study. The optimized synthesis conditions retained for the samples measured either by XRPD and Neutron Powder Diffraction were 2.5 GPa and 620°C in the Conac anvil. For the TEM study in addition to this sample (S1) a mixture of 6H-B and 3C obtained at 3.7GPa and 620°C (S2) was studied.

## 2.2. X-Ray and neutron powder diffraction

Sample S1 was characterized by x-ray powder diffraction (XRPD) using a Bruker D8 diffractometer with CuK$\alpha$1 radiation (1.5406 Å) selected by a Ge (111) primary beam monochromator in the 2$\Theta$ range 10 ~ 90° with a 0.01° step size. No impurity phase could be detected. A neutron powder diffraction (NPD) experiment was carried out on the two-axis diffractometer D1B at Institut Laue Langevin (ILL). The NPD pattern was recorded at 300 K with a wavelength of 1.28 Å selected by the (311) Bragg reflection of a germanium monochromator. The XRPD and NPD data were used to perform a combined Rietveld refinement of the crystal structure in order to check the space group symmetry and the substitutional disorder. The Fullprof Suite programs were used to carry out the refinements and analyze the results [15].

## 2.3. Transmission Electron Microscopy

Specimen for transmission electron microscopy were prepared by crushing a small amount of the sample in an agate mortar containing ethanol. A droplet of the suspension was deposited on a copper grid covered by a holey carbon film.

Z-contrast images were acquired with a probe corrected JEOL ARM 200 microscope equipped with a scanning transmission electron microscopy (STEM) unit and a high-angle annular dark-field (HAADF) detector with a point resolution of 0.078 nm. On such images, contrast is proportional to the sum of $Z^n$ of the elements present in the projected atomic columns. The exponent $n$ varies from 2 (Rutherford scattering) to 1.5 (thermal diffuse scattering), depending on the experimental conditions such as angles on the HAADF detector or crystal thickness. Consequently, the more an atomic column contains heavy atoms, the brighter its projection appears. In 6H-B $Ba_3NiSb_2O_9$, we therefore expect the projections of Ba ($Z$ = 56) and Sb ($Z$ = 53) atomic columns to appear as bright dots, while those of Ni ($Z$ = 28) should be less bright and O ($Z$ = 8) columns dark. Measurements on the image were carried out using the software Gatan Digital Micrograph.

An electron diffraction study of $Ba_3NiSb_2O_9$ was conducted on a Philips CM300ST transmission electron microscope equipped with a Nanomegas Spinning Star precession unit. Several particles of two samples (S1 and S2) were studied by electron diffraction in zone axis geometry and from each particle diffraction patterns from several zone axes were recorded on a TVIPS F416 CMOS camera. Since the diffracted intensities of classical selected area electron diffraction (SAED) patterns are not reliable because of multiple diffraction effects, additionally to the SAED patterns, we recorded precession electron diffraction (PED) [16] patterns with a precession angle of 2.5°. In PED patterns multiple diffraction is largely reduced and the resulting intensities can be used for the determination of symmetry elements in the diffraction patterns.

## 3. Results and discussion

### 3.1. Joint XRPD and NPD structure refinement

The main goal of the XRPD-NPD experiment was to determine the presence of structural disorder in the cation stacking sequence of the 6H-B samples. Due to the difference in scattering contrasts between atomic species for the two radiations (for neutrons, Ni and O are stronger scatterers than Ba and Sb), combined refinements using both types of data are expected to yield more accurate results. Since such disorder would be reflected in the symmetry of the compound, joint refinements were carried out exactly in the same way for space groups *P6$_3$mc* and *P6$_3$/mmc*. The background for NPD was treated as a polynomial; for XRPD it was interpolated from selected points in the patterns. The profile parameters were refined independently for the two data sets, using the Thomson-Cox-Hasting description of the pseudo-Voigt function [17]. The Bérar-Baldinozzi approximation was used to take account of low angle peak asymmetry [18]. For the final refinement, the positional parameters and isotropic individual atomic displacement parameters of all atoms were refined together with cell, profile and global parameters. Since the *P6₃mc* space group has no fixed origin along the **c** axis, the Ba1 atom was fixed at the (0 0 1/4) position. The occupancies of the Sb/Ni sites corresponding to the face sharing octahedra were refined with the constraint that each site should be fully occupied. This resulted in 18 global and profile parameters, plus 22 and 13 intensity affecting parameters for *P6₃mc* and *P6₃/mmc*, respectively. The weights of the XRPD and NPD data collections were fixed at ½. Attempts to modify this value to ¼/¾ or ¾/¼ did not lead to any substantial modification of the results. The refinement results for both space groups are listed in table 1, while the main interatomic distances and bond valence sums (BVS) calculated using the BondStr [19] program, are given in table 2. The Rietveld plots for both radiations are displayed in figure 2.

The refinement in the non-centrosymmetric space group *P6₃mc* did not lead to any improvement of the agreement factors, which were in fact slightly worse than for centrosymmetric *P6₃/mmc*. As stated above, both structures differ only by the presence of a mirror plane at $z = 1/4$ for *P6₃/mmc* which is absent for *P6₃mc*. Choosing the latter space group would only be justified if any sizeable discrepancies between positions or occupancies of atoms related by this mirror plane in the high symmetry group are found in the refinement using the lower symmetry. The main difference would be an ordering between $Ni^{2+}$ and $Sb^{5+}$ cations between sites 1a and 1b forming the octahedra dimers, which is not observed since both cations appear from the refinement as quasi-evenly distributed on both sites. Attempts to force such order in the refinement systematically yielded much poorer agreement and freeing the occupancies always led back to the same even distribution. Therefore, the choice of the non-centrosymmetric space group *P6₃mc* is not justified from the XRPD and NPD data.

We can try to argue about the nature of the cation disorder from crystal-chemical considerations. In the 6H-A structure, the inter-cationic distance through the shared octahedron face reported by Jacobson *et al.* is 2.84(1) Å [14], while it is only 2.679(2) Å for the 6H-B structure, as found by Köhl *et al.* [8]. The lower value for the 6H-B phase can be explained by the smaller electrostatic repulsion between the $Ni^{2+}$-$Sb^{5+}$ cations (6H-B) than between a pair of $Sb^{5+}$ cations (6H-A). In the present refinement this distance is 2.72(3) Å, close to the value of Köhl *et al*. It is therefore probable that the face sharing octahedra are occupied by couples of $Sb^{5+}/Ni^{2+}$ cations, but not by $Sb^{5+}/Sb^{5+}$ nor $Ni^{2+}/Ni^{2+}$ couples. However, from these data alone, it is not reasonable to go beyond this statement: our observations are compatible with a stacking disorder along the **c**-axis as well as a disorder between ($Ni^{2+}/Sb^{5+}$) and ($Sb^{5+}/Ni^{2+}$) dimers in the double layers. The consequences are quite important for the physical properties, since in the latter case the frustrated magnetic $Ni^{2+}$ triangular planes would no longer exist. The investigations using electron diffraction and Z-contrast imaging reported below principally aimed at solving this issue.

### 3.2. TEM Z-contrast imaging

The Z-contrast image projected along b-axis is displayed on figure 3a, with the corresponding projection of the 6H-B structure on figure 3b. On the image lines composed of 7 bright dots linked to each other in a zig-zag configuration are observed that correspond on the structure projection to regular alternations of Ba with either Ni or Sb atoms. On the image the dark zig-zag lines in between correspond to the O atom projections. A typical series of such bright dots are labelled on the image as 1, 2, 3, 4, 5, 6, 7 on a first line and a, b, c, d, e, f, g on the following line, and the corresponding atom projections are visible in figure 3b. Intensity profiles obtained on both lines are displayed on figure 3c. Intensity profiles on equivalent rows have been checked to present the same features. These experimental profiles are compared to the theoretical ones presented on figure 3d that would be obtained from the 6HB structural model by calculating the maxima of peaks for each atom projections considering $I \propto Z^n$, with $n = 1.7$ (typical averaged value). The calculated intensities were scaled to fit the intensities of the Ba positions. Note that the positions 2, b and 6, f correspond to the pairs of cations occupying a single dimer of face-sharing octahedra. We observe on the first profile that intensities 1, 3, 4, 5, 7 are strong, as well as intensities a, c, d, e, g on the second profile. The comparison to the theoretical intensities from the 6H-B structure shows a good agreement to the Ba, Ba, Sb, Ba, Ba atom projections respectively. The experimental intensities at positions 2, 6, b and f however are significantly different from the theoretical ones. On positions 6 and b we expect higher intensities corresponding to Sb atom projections, similar to positions 1, 3, 4, 5, 7 or a, c, d, e, g. On the contrary, on positions 2 and f we expect much lower intensities corresponding to Ni atom columns. The experimental intensities of these 4 columns are almost equal and intermediate to what would be expected from pure Sb or pure Ni atom columns. A possible interpretation of this result is that throughout the thickness of the sample (between 10 to 50 nm) the columns contain roughly the same amount of Ni and Sb atoms. Consequently, since the positions 6 or f and 2 or b are those forming the dimers of face-sharing octahedra, this corroborates the cationic disorder observed on these sites by the joint XRPD and NPD refinements reported above. Furthermore our Z-contrast imaging observations involve the existence of a Ni/Sb substitution along the **b**-axis, i.e. within each octahedral layer forming the dimers (see Fig. 1). Since we are dealing with a **b**-axis projection of the structure, it is not possible to discriminate between a Ni/Sb order along this b-axis direction, or an alternation of nano-size domains of Ni or Sb atoms, or a complete random disorder between these two species.

### 3.3. TEM electron diffraction

Electron diffraction offers the possibility of obtaining single crystal diffraction data from very small volumes of a sample. We can therefore avoid peak overlap and compared to XRPD and NPD, PED can provide insight on the structure at a much more local level.

All of the electron diffraction patterns could be indexed with the hexagonal unit cell reported in the literature for the 6H-B phase of $Ba_3NiSb_2O_9$ ($a$ = 5.8 Å, $c$ = 14.3 Å [8,9]). However, the PED patterns obtained from very thin particles showed intensity distributions lacking a mirror symmetry perpendicular to the **c*** direction (figure 4a). This is incompatible with a 6-fold rotation axis along **c*** since the 6-fold axis combined to Friedel's law would lead to a *2mm* symmetry of the [0 1 0] zone axis, whereas figure 4a clearly shows only a symmetry *2* without any mirror. This optical impression is corroborated by the profile (figure 4c) obtained from the boxed area in figure 4a. Not only the 2 0 4 reflection is much more intense than the 2 0 -4, but the relative intensities of the other reflections are significantly different for positive and negative *l* values. Therefore the crystal structure cannot have hexagonal symmetry but is at most trigonal.

The PED patterns of the [0 1 0] zone axis obtained from other particles didn't systematically show such a distinctive breach of the mirror symmetry perpendicular to **c***.

Figure 4b shows one such example obtained from a particle that was less thin than the one yielding the PED pattern of figure 4a. The profile shown in figure 4d, which was obtained from the boxed area in figure 4b, still shows a slight asymmetry, but it is much less pronounced than in the previous case. This leveling of the relative intensities could either be explained by multiple diffraction effects or by the superposition of different orientations of the same diffraction pattern as in figure 4a.

However, looking at the 00*l* row of reflections can help distinguish between the two interpretations. This row is symmetric with respect to the transmitted beam due to Friedel's law. Therefore in different orientations of the same diffraction pattern odd reflections will superimpose with odd and even with even reflections. The characteristic weakness of the odd with respect to the even reflections would therefore not be altered by such a superposition.

In figure 4b one can still see very distinct intensity differences between the reflections with odd or even values of *l*. Comparing this to the 00*l* row of figure 4a shows that there are no major differences, which would indicate that dynamical scattering is not predominant in the diffraction pattern of figure 4b and we can conclude that the differences between figures 4a and 4b stem from the superposition of different domains.

The superposition of different domains in the thicker particle yields an intensity distribution that (almost) restores the mirror symmetry, while in the thinner samples one orientation is dominant in the diffracting volume and the absence of the mirror perpendicular to **c\*** is clearly observed.

In order to determine the space group of the crystal the systematic extinctions of reflections were studied in PED patterns. No extinctions for general *hkl* reflections were observed indicating a hexagonal and not a rhombohedral lattice for the trigonal space group. The additional extinction conditions can be observed on the [1 -1 0] zone axis (figure 4e). In this zone axis all reflections are of type *hhl* and all of the high intensity reflections have even values for *l*. A systematic absence of all *hhl* reflections with odd *l* would indicate the existence of a **c** glide plane. However, on closer inspection one can also observe *hhl* reflections with odd *l*. Note that multiple diffraction can't bring intensity to extinct reflections in this zone axis, since a combination of *hhl* reflections with even *l* can't yield an *hhl* reflection with odd *l*. The observation of reflections, 003, 111, 112 and 113 amongst others therefore shows that there is no systematic extinction for *hhl* reflections. In addition, the rather strong 003 reflection, in addition to the 00*l* reflections with even *l*, shows that there is no extinction for 00*l* reflections. We can therefore conclude that the extinction symbol is *P---* corresponding to a primitive trigonal unit cell without any glide planes nor screw axes.

## 3.4. Discussion on the structure

The electron microscopy observations indicate that the actual symmetry would be at most trigonal, since the observed reflections do not respect neither the hexagonal symmetry (the $6_3$ axis disappears) nor the **c** glide plane. A threefold screw axis could also be excluded, which leads to several possible space groups, the highest symmetry of which is *P3m1*. Here we consider the space group *P3m1* rather than P-*3m1* since the latter is not a subgroup of *P6$_3$mc*, the space group of the 6H-B phase, which is regarded to be the parent phase of the one we describe here. We did try to refine the structure using the same XRPD and NPD data as above in space group *P3m1* and *P3*. These refinements were strongly impeded by parameter correlations, especially for *P3*. The final agreements were similar to those reported above for *P6$_3$/mmc* and *P6$_3$mc* and the refinements did not show any sizeable improvement or structural modification (for example cation ordering on the dimer sites) which could justify the symmetry lowering based on these data. Therefore, from the point of view of powder diffraction, one can consider the observed space group as being *P6$_3$/mmc*., with a complete disorder between $Sb^{5+}$ and $Ni^{2+}$ on the face-sharing octahedral sites forming the dimers. The electron microscopy experiments help us to get a more detailed picture of the structure. The lowering of the local symmetry to *P3m1* (or lower) demonstrates that the stacking order of type -A($Sb^{5+}$)-A($Sb^{5+}$)=A($M^{2+}$)-A($Sb^{5+}$)-A($Sb^{5+}$)=A($M^{2+}$)-A($Sb^{5+}$)- between the cation layers which is directly related to the presence of the $6_3$ axis and **c** glide plane in the 6H-B structure can't be maintained. For XRPD and NPD, the stacking sequences -A($Sb^{5+}$)-A($Sb^{5+}$)=A($M^{2+}$)-A($Sb^{5+}$)- and -A($Sb^{5+}$)-A($M^{2+}$)=A($Sb^{5+}$)-A($Sb^{5+}$)- appear as equiprobable and are randomly distributed along the **c**-axis. The observations from electron diffraction show that there must be an order of the cation layers, but one that isn't compatible with neither a $6_3$-axis nor a **c**-glide plane, while keeping the same periodicity of the unit cell. One such order is proposed in figure 5a where the cation occupation of every second dimer is inversed. The stacking order of cation planes then is -A($Sb^{5+}$)-A($Sb^{5+}$)=A($M^{2+}$)-A($Sb^{5+}$)-A($M^{2+}$)=A($Sb^{5+}$)-

A($Sb^{5+}$)-. In this case, the A($M^{2+}$) layers are separated alternately by one and three A($Sb^{5+}$) layers, while in the 6H-A and 6H-B stacking, the separation is always made of two such layers. Hence the disappearance of the $6_3$-axis and **c**-glide plane.

The simulated precession electron diffraction pattern of zone axis [0 1 0] obtained for a precession angle of 2.5° and a thickness of 10 nm from this model is shown in figure 5b. The simulation clearly shows the absence of the mirror symmetry perpendicular to the **c\*** axis. Note the close resemblance of the intensities in the 00l row between experiment and simulation corroborating our model with respect to the stacking of the cation layers along the **c** axis. In particular the high intensity of the 003 reflection is in agreement with the experimental diffraction pattern of figures 4a and e. This reflection is extinct in the $P6_3mc$ space group of 6H-B.

A possible reason that the simulation does not match perfectly the experimental diffraction pattern in the other systematic rows is that atomic displacements relative to the 6H-B structure have not been taken into account in the model. Only the occupation of some Ni and Sb sites has been modified relative to the 6H-B structure.

The fact that this symmetry lowering can be observed by electron diffraction, but not by XRPD and NPD, indicates that ordered domains of this sequence have some sizeable extension. The Z-contrast imaging results reveal that this extension is not infinite, with domains being superimposed in a projection along the **b** axis of the structure. From the typical coherence lengths of XRPD and NPD, and the size of the area sampled by electron microscopy, the domain size can be roughly estimated between 1 and 10 nm. The change of a A($Sb^{5+}$)=A($M^{2+}$) dimer to a A($M^{2+}$)=A($Sb^{5+}$) either in the **c**-direction or in the **a**- or **b**-directions would lead to a stacking fault which could represent the limit between neighboring domains.

For the interpretation of the physical property measurements, the resulting model would correspond to planar domains of $Ni^{2+}$ cations forming a triangular lattice, with a typical domain size of a few nm in the **ab**-plane, stacked with the -A($Sb^{5+}$)-A($Sb^{5+}$)=A($M^{2+}$)-A($Sb^{5+}$)-A($M^{2+}$)=A($Sb^{5+}$)-A($Sb^{5+}$)- sequence in the **c**-direction (as shown in figure 5) over a few unit cell distances.

# 4. Conclusion

In the search for an experimental realization of a quantum spin liquid, $Ba_3NiSb_2O_9$ with a 6H structure appears as a promising candidate. In order to ascertain the interpretation of the physical behavior observed for this compound, complex issues concerning its structure and the presence of faults or domains in the triangular layers of magnetic cations had to be sorted out. We performed high temperature/ high pressure synthesis aiming at the synthesis of the 6H-B form and obtained a compound that shows no sign of magnetic ordering at low temperature as expected for QSL behaviour. However, we evidenced that our sample displays a substantially disordered structure compared to what is reported in the literature [9]. In such a sample, the absence of magnetic ordering could be simply related to the structural disorder of the magnetic cation framework, in which the triangular planes are not preserved. This would lead to a system with much less physical interest. We performed a thorough investigation of the structure at various scales using a combination of x-ray, neutron and electron diffraction techniques and electron Z-contrast imaging. Electron diffraction on a thin particle allowed showing that the symmetry of the structure is not hexagonal but trigonal. Ni-triangle planes can effectively be present in this structure but only in domains of up to 10 nm in size. . At the larger scale seen by x-rays and neutrons, typically several 10s nm we observe an average of several domains and the symmetry becomes hexagonal with a statistical Ni/Sb disorder on the cation sites of the face sharing octahedra. Based on this work, we propose a model for the local stacking arrangement of the octahedron layers which is consistent with all our observations. In this model, the sequence of Ni and Sb containing layers is different from those reported for the 6H-A and 6H-B forms. This arrangement appears as well ordered over distances extending between 1 and 10 nm, within which the triangular arrangement of the magnetic cations is preserved.

# Acknowledgements

The syntheses were performed thanks to the help of Céline Goujon and Murielle Legendre from Institut Neel's technical group "Instrumentation". We thank CRG-D1B at ILL for the technical support. We thank Jeffrey Quilliam, Fabrice Bert, Philippe Mendels for fruitful discussions. This work was supported by the French Agence Nationale de la Recherche under Grant "SPINLIQ" No. ANR-12-BS04-0021.# References

[1] P. W. Anderson Mater Res Bull 8, (1973), 153-160
[2] L. Balents, Nature, 464, (2010), 199-208
[3] H. D. Zhou, E. S. Choi, G. Li, L. Balicas, C. R. Wiebe, Y. Qiu, J. R. D. Copley, and J. S. Gardner, Phys Rev Lett 106, (2011), 147204
[4] S. Nakatsuji, K. Kuga, K. Kimura, R. Satake, N. Katayama, E. Nishibori, H. Sawa, R. Ishii,
M. Hagiwara, F. Bridges, T. U. Ito, W. Higemoto, Y. Karaki, M. Halim, A. A. Nugroho,
J. A. Rodriguez-Rivera, M. A. Green, C. Broholm, Science 336, (2012), 559-563
[5] J. A. Quilliam, F. Bert, E. Kermarrec, C. Payen, C. Guillot-Deudon, P. Bonville, C. Baines, H. Luetkens, and P. Mendels, Phys Rev lett, 109, (2012), 117203
[6] H. D. Zhou, Cenke Xu, A. M. Hallas, H. J. Silverstein, C. R. Wiebe, I. Umegaki, J. Q. Yan, T. P. Murphy, J.-H. Park, Y. Qiu, J. R. D. Copley, J. S. Gardner, and Y. Takano
Phys Rev Lett 109, (2012), 267206
[7] G. Blasse, J. Inorg. Nucl. Chem.27, (1965), 993-1003
[8] P. Köhl, D. Reinen, Z. Anorg. Allg. Chem. 433 (1977) 81-93
[9] P.D. Battle, C. W. Jones, P. Lightfoot, R. Strange, J. Solid State Chem. 85, (1990), 144-150
[10] J. G. Cheng, G. Li, L. Balicas, J. S. Zhou, J. B. Goodenough, Cenke Xu, and H. D. Zhou, Phys. Rev. Lett., 107, (2011), 197204
[11] Y. Doi, Y. Hinatsu, K. Ohoyama J. Phys.: Condens.Matter 16 (2004) 8923-8935
[12] J. A. Quilliam, F. Bert, A. Manseau, C. Darie, C. Deudon, C. Payen, C. Baines, A. Amato, P. Mendels to be published
[13] P. D. Battle, C. M. Chin, S. I. Evers, M. Westwood, J. Solid State Chem. 227 (2015) 1-4
[14] A.J. Jacobson and A.J. Calvert, J. Inorg. Nucl. Chem., 40, (1978), 447-449
[15] J. Rodriguez-Carvajal, Physica B, 192, (1992), 55-69.
[16] R. Vincent, P.A. Midgley, Ultramicroscopy, 53,(1994), 271-282
[17] P. Thompson, D.E. Cox, J.B. Hastings, J. Appl. Cryst., 20, (1987), 79-83.
[18] J.F. Bérar and G. Baldinozzi, J. Appl. Cryst., 26, (1993), 128-129.
[19] J. Rodrıguez-Carvajal, BondStr, 2010, www.ill.eu/sites/fullprof/

Table 1: Refined structural parameters and agreement factors from the joint Rietveld refinement of NPD + XRPD data for 6H-B $Ba_3NiSb_2O_9$.

| Atom | Wyck. | S.O.F. | x/a | y/b | z/c | U [Å$^2$] |
|---|---|---|---|---|---|---|
| Ba1 | 2b | 1.0 | 0 | 0 | 1/4 | 0.0038(7) |
| Ba2 | 4f | 1.0 | 1/3 | 2/3 | 0.9009(2) | 0.0126(8) |
| Sb1 | 4f | 0.504(6) | 1/3 | 2/3 | 0.1549(1) | 0.0057(5) |
| Ni1 | 4f | 0.496 | 1/3 | 2/3 | 0.1549 | 0.0057 |
| Sb2 | 2a | 1.0 | 0 | 0 | 0 | 0.0026(7) |
| O1 | 6h | 1.0 | 0.4893(5) | 0.9785(5) | 1/4 | 0.0130(5) |
| O2 | 12k | 1.0 | 0.1624(5) | 0.3247(5) | 0.4206(1) | 0.0126(4) |

Space group $P6_3/mmc$, a= 5.7926(2), c= 14.2841(6)
XRPD: Rp: 15.8, Rwp: 20.6, Chi2: 1.66 ; $R_{Bragg}$: 2.64
NPD: Rp: 7.18, Rwp: 6.77, Chi2: 71.7

| Atom | Wyck. | S.O.F. | x/a | y/b | z/c | U [Å$^2$] |
|---|---|---|---|---|---|---|
| Ba1 | 2a | 1.0 | 0 | 0 | 1/4 | 0.002(1) |
| Ba2 | 2b | 1.0 | 1/3 | 2/3 | 0.103(2) | 0.005(4) |
| Ba3 | 2b | 1.0 | 1/3 | 2/3 | 0.404(2) | 0.023(6) |
| Ni1a | 2b | 0.48(8) | 1/3 | 2/3 | 0.658(2) | 0.005(3) |
| Sb1a | 2b | 0.52 | 1/3 | 2/3 | 0.658 | 0.005 |
| Sb2 | 2a | 1.0 | 0 | 0 | 0.506(1) | 0.0019(7) |
| Sb1b | 2b | 0.47(8) | 1/3 | 2/3 | 0.848(2) | 0.007(3) |
| Ni1b | 2b | 0.54 | 1/3 | 2/3 | 0.848 | 0.007 |
| O1 | 6c | 1.0 | 0.161(1) | 0.839(1) | 0.586(2) | 0.015(3) |
| O2 | 6c | 1.0 | 0.5108(3) | 0.4893(3) | 0.258(1) | 0.0121(7) |
| O3 | 6c | 1.0 | 0.836(1) | 0.164(1) | 0.427(1) | 0.010(3) |

Space group $P6_3mc$, a= 5.7926(2), c= 14.2842(6)
XRPD: Rp: 18.0, Rwp: 22.6, Chi2: 1.63 ; $R_{Bragg}$: 2.77
NPD: Rp: 7.25, Rwp: 6.84, Chi2: 72.8

Table 2: principal interatomic distances (Å) and bond valence sums (v.u;) for the Rietveld refinement of 6H-B $Ba_3NiSb_2O_9$ with space group $P6_3/mmc$.

| | | |
|---|---|---|
| Ba1-O1 x2: 2.898(3) | Ba2-O1 x1: 2.796(3) | Sb1/Ni1-O1 x2: 2.072(2) |
| Ba1-O1 x2: 2.899(3) | Ba2-O1 x2: 2.795(3) | Sb1/Ni1-O1 x1: 2.072(3) |
| Ba1-O2 x2: 2.931(2) | Ba2-O2 x1: 2.910(2) | Sb1/Ni1-O2 x1: 2.026(3) |
| Ba1-O2 x2: 2.931(2) | Ba2-O2 x1: 2.911(2) | Sb1/Ni1-O2 x2: 2.026(3) |
| Ba1-O2 x2: 2.931(2) | Ba2-O2 x1: 2.910(3) | |
| Ba1-O2 x2: 2.931(2) | Ba2-O2 x1: 2.910(3) | Average: 2.049(1) |
| | Ba2-O2 x1: 2.911(3) | BVS for $Sb^{5+}$: 4.50(1) |
| Average: 2.915 | Ba2-O2 x1: 2.910(3) | BVS for $Ni^{2+}$: 2.067(6) |
| BVS for $Ba^{2+}$: 2.191(4) | Ba2-O2 x1: 3.073(3) | |
| | Ba2-O2 x1: 3.073(3) | Sb2-O2 x2: 1.985(3) |
| | Ba2-O2 x1: 3.073(3) | Sb2-O2 x4: 1.985(2) |
| | | |
| | Average: 2.9223(9) | Average: 1.985(1) |
| | BVS for $Ba^{2+}$: 2.219(5) | BVS for $Sb^{5+}$: 5.34(2) |
| | | BVS for $Ni^{2+}$: 2.453(7) |

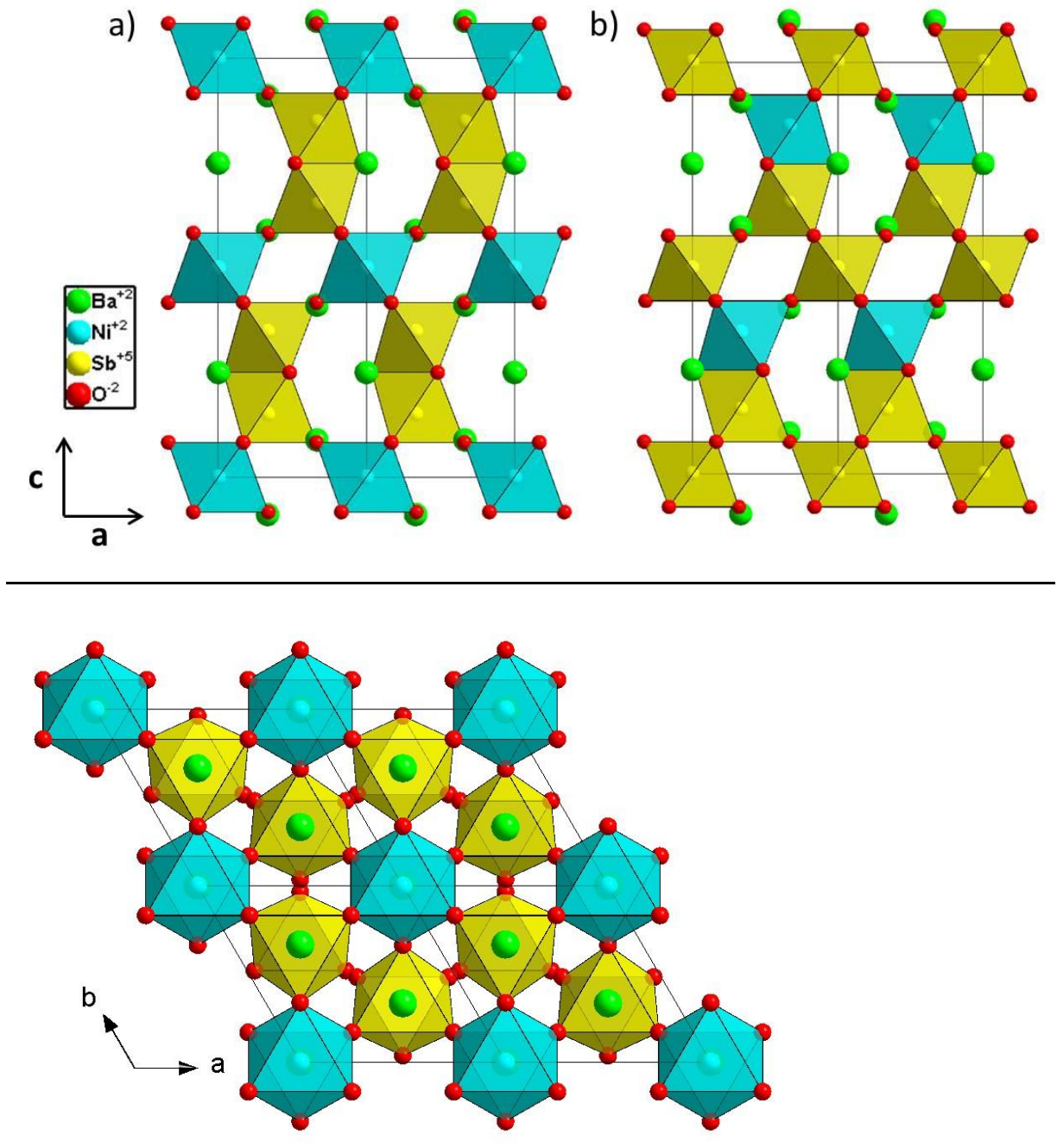

Figure 1: Structures of (a) 6H-A-Ba$_3$NiSb$_2$O$_9$ and (b) 6H-B-Ba$_3$NiSb$_2$O$_9$. Projection along **b** axis. (c) projection along **c** axis showing the Ni triangular planes.

a)
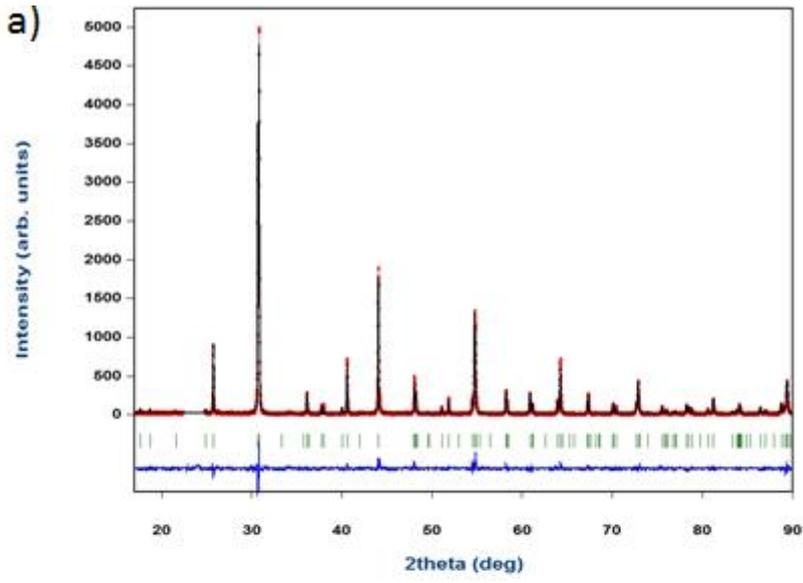

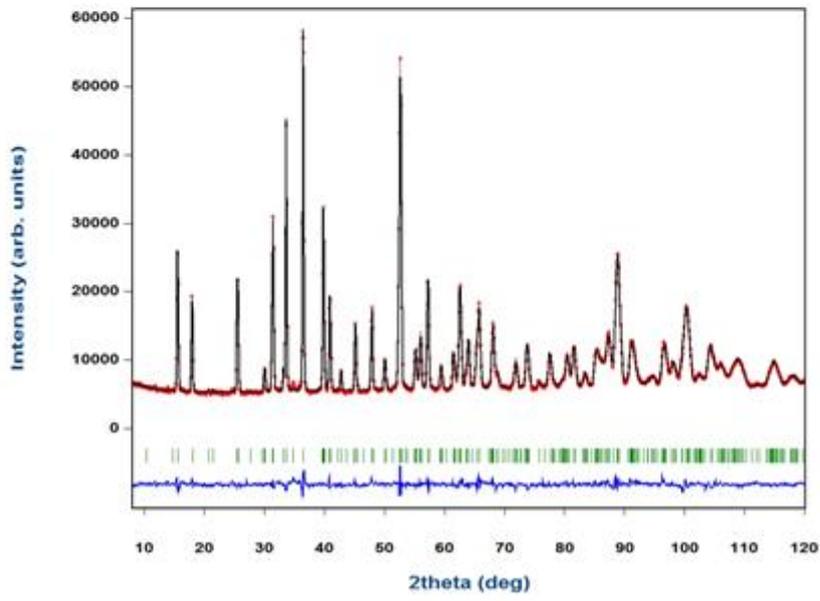

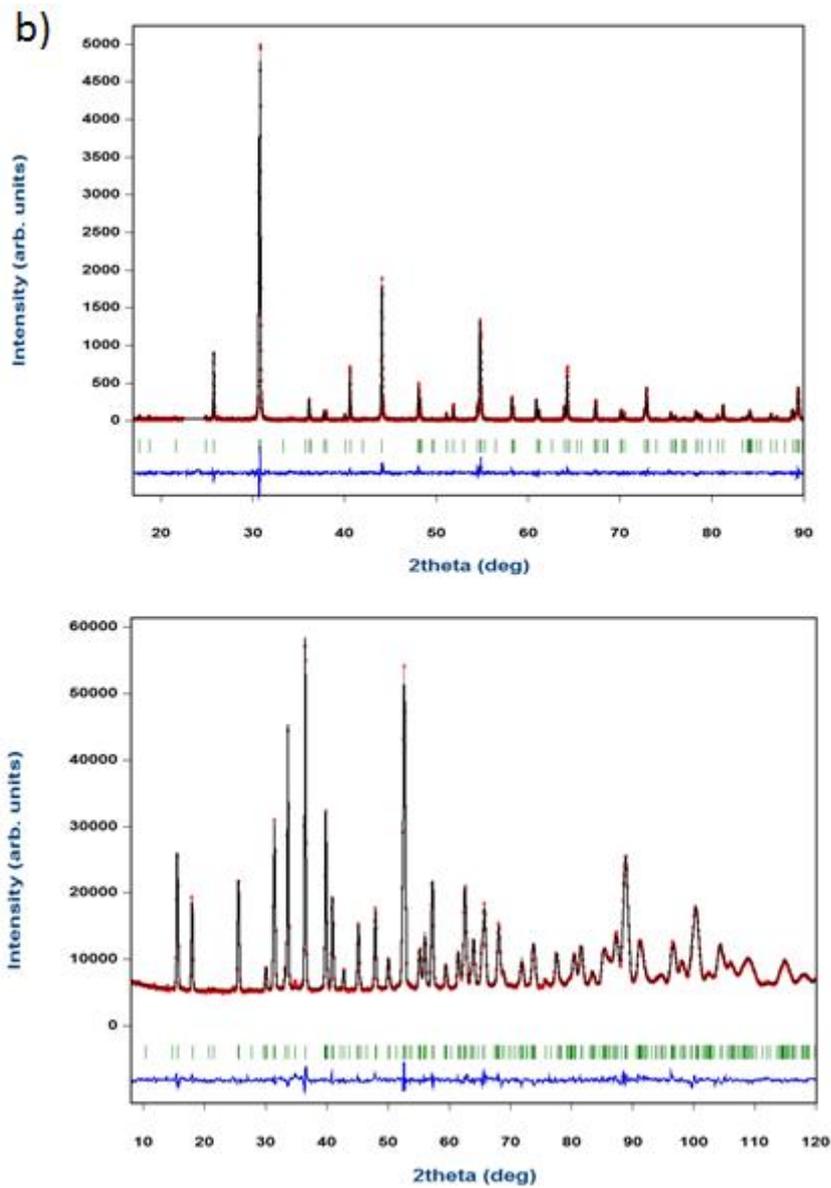

Figure 2: (a) XRD (top panel) and NPD (bottom panel) refinements of 6HB-Ba$_3$NiSb$_2$O$_9$ performed with P6$_3$/mmc at 295K (b) XRD (top panel) and NPD (bottom panel) refinements of 6HB-Ba$_3$NiSb$_2$O$_9$ performed with P6$_3$mc at 295K. Observed (red), calculated (black) and difference (blue) plots are shown, and Bragg reflections are indicated by green tick marks.

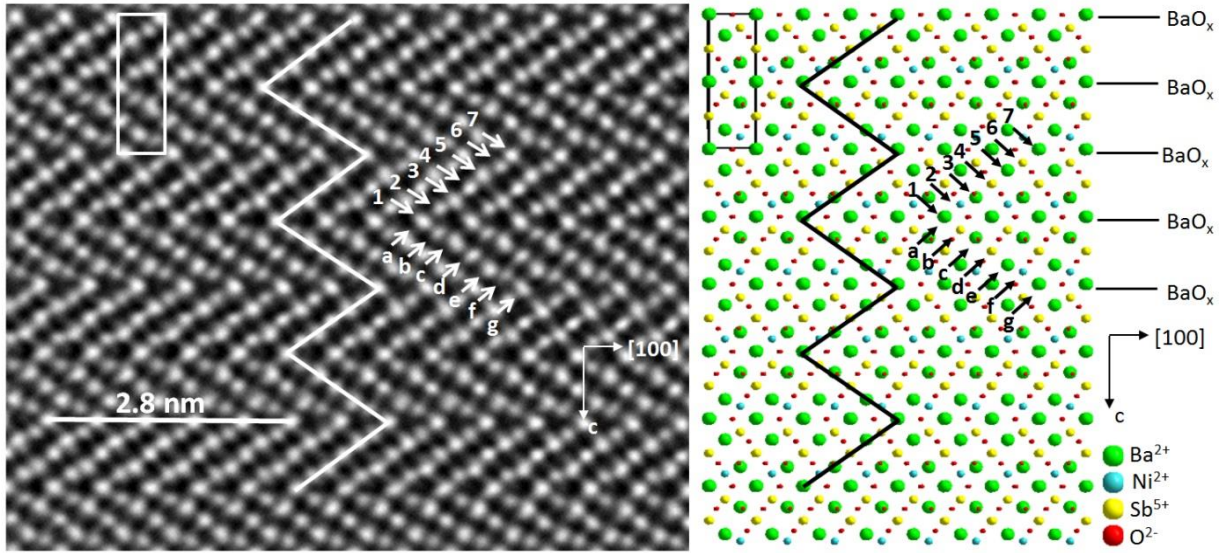

(a)                                 (b)

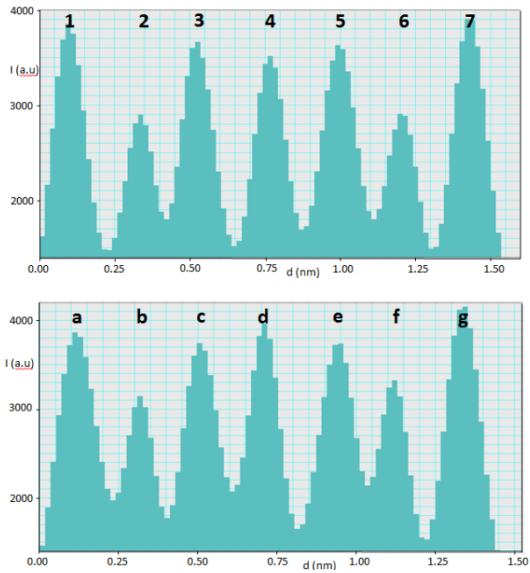
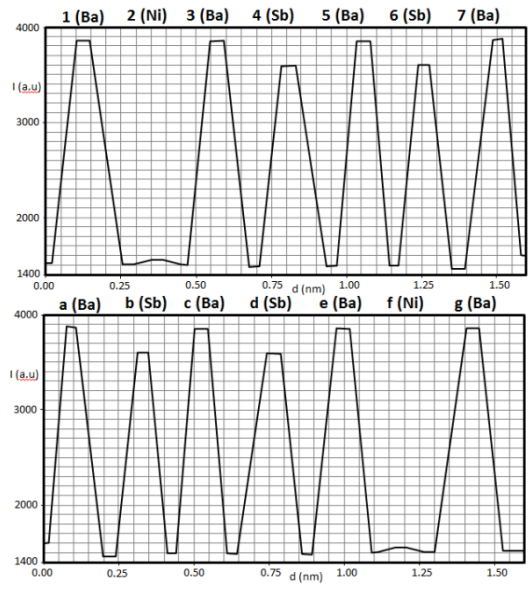

(c)                                 (d)

Figure 3: (a) Z-contrast image of $Ba_3NiSb_2O_9$ along its **b** axis, (b): corresponding projection of the 6H-B structure, (c) intensity profiles obtained from the Z-contrast image, (d) theoretical intensity profiles from 6H-B structure.

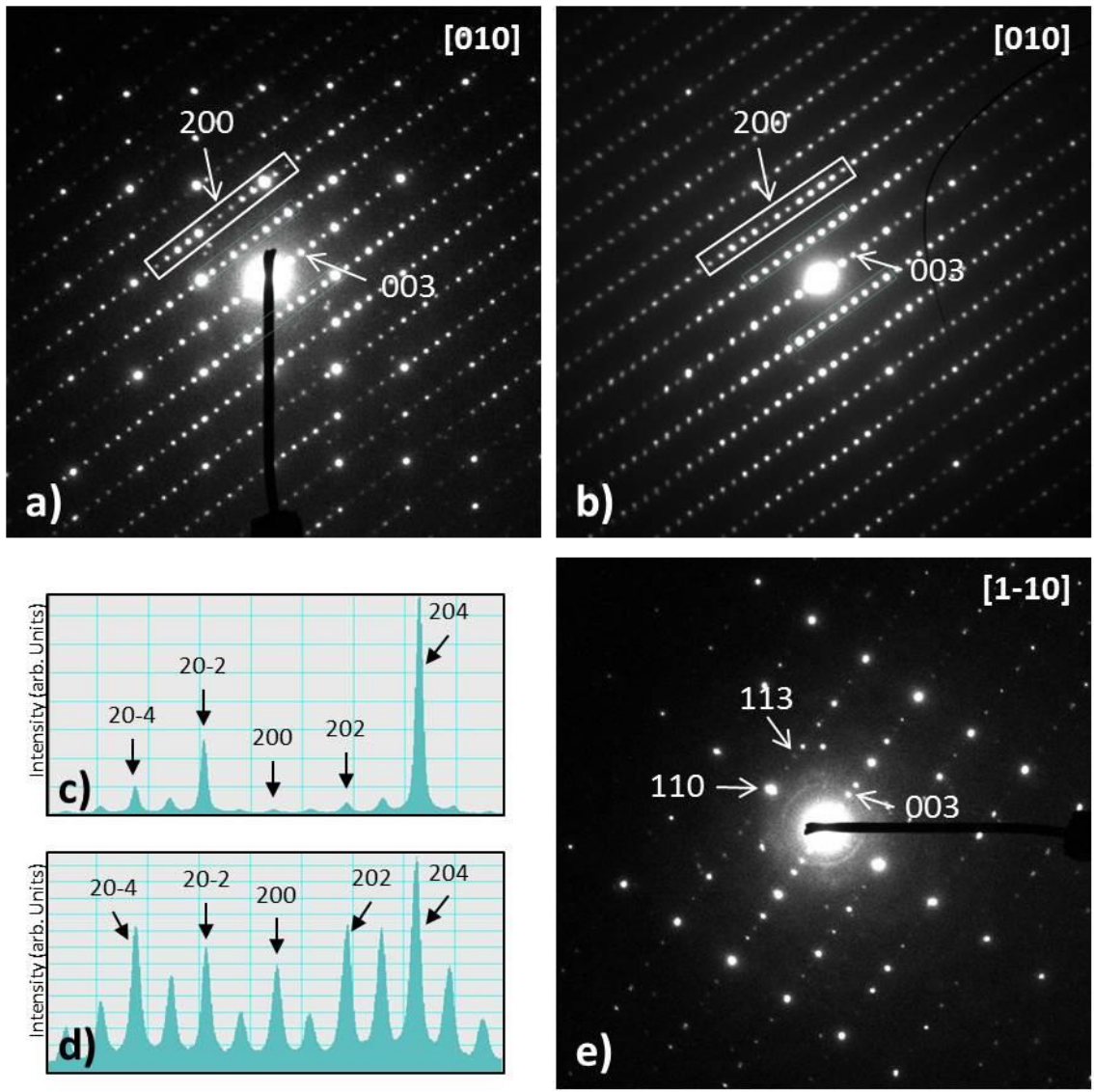

Figure 4: Typical precession electron diffraction patterns of 6H-B-$Ba_3NiSb_2O_9$ along [010] on (a) a very thin particle and (b) a thicker particle. (c,d) Intensity profiles related to boxed areas on (a,b) respectively, clearly showing the absence of a mirror symmetry. (e) [1 -1 0] PED zone axis, evidencing the absence of systematic extinction for *hhl* reflections.

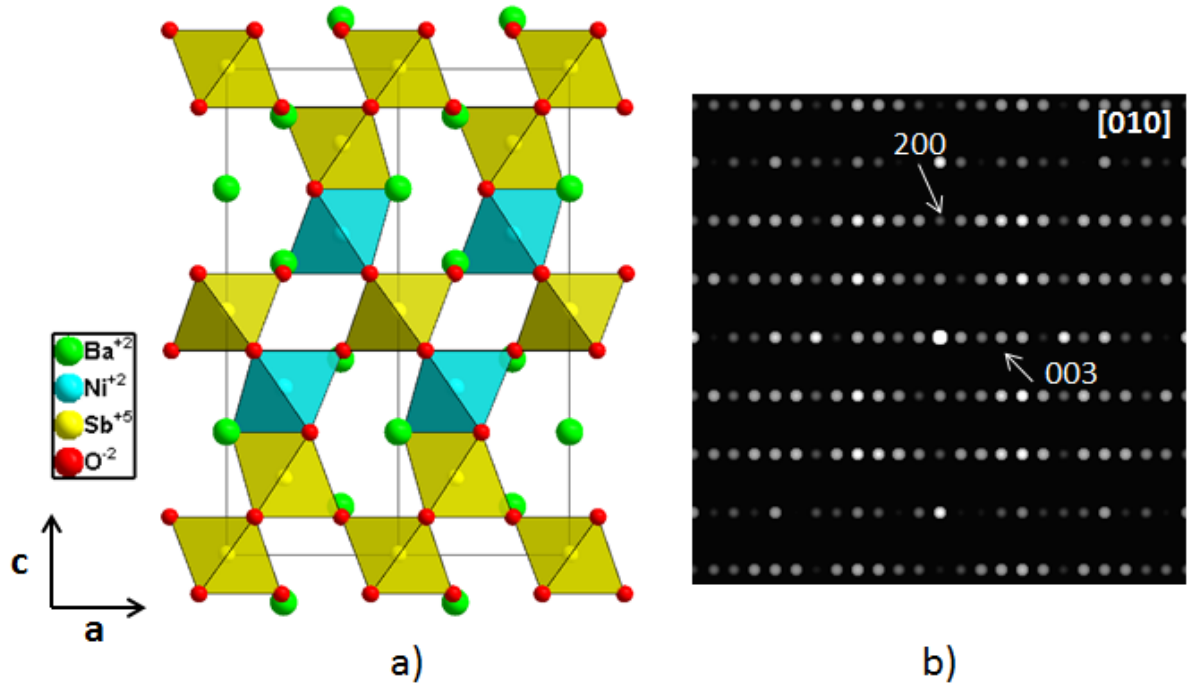

Figure 5:

(a) Proposed structural model in space group *P3* projected along the **b** axis compatible with our results.

(b) Corresponding simulated precession electron diffraction pattern of the zone axis [0 1 0] (Precession angle = 2.5°, sample thickness = 10 nm) evidencing the absence of mirror symmetry perpendicular to **c***.